\documentclass[preprint,5p,twocolumn]{elsarticle}
\usepackage{hyperref}
\usepackage{booktabs}
\usepackage{amsmath}

\title{A Test of Bolometric Properties of Tm-containing Crystals as a Perspective Detector for the Solar Axion Search.}

\author[1]{E.~Bertoldo}
\author[2]{A.~V.~Derbin}
\author[2]{I.~S.~Drachnev}
\author[3]{M.~Laubenstein}
\author[4]{D.~A.~Lis}
\author[1]{M.~Mancuso}
\author[2]{V.~N.~Muratova}
\author[5]{S.~Nagorny}
\author[3]{S.~Nisi}
\author[1]{F.~Petricca}
\author[6]{V.~V.~Ryabchenkov}
\author[6]{S.~E.~Sarkisov}
\author[2]{D.~A.~Semenov}
\author[4]{K.~A.~Subbotin}
\author[2]{E.~V.~Unzhakov}
\author[4]{E.~V.~Zharikov}

\address[1]{Max-Planck-Institut f\"ur Physik, 80805 M\"unchen, Germany}
\address[2]{NRC Kurchatov Institute, Petersburg Nuclear Physics Institute, 188309 Gatchina, Russia}
\address[3]{INFN, Laboratori Nazionali del Gran Sasso, 67010 Assergi, Italy}
\address[4]{Prokhorov General Physics Institute of the Russian Academy of Sciences, 119991 Moscow, Russia}
\address[5]{Queen’s University, Physics Department, K7L 3N6 Kingston, Canada}
\address[6]{NRC Kurchatov Institute, 123182 Moscow, Russia}

\newcommand{\TmAlO}{$\mathrm{Tm}_{3}\mathrm{Al}_{5}\mathrm{O}_{12}$}  
\newcommand{\YON}{$\mathrm{Y}_{2}\mathrm{O}_{3}(4\mathrm{N})$}  
\newcommand{\TmON}{$\mathrm{Tm}_{2}\mathrm{O}_{3}(5\mathrm{N})$}  
\newcommand{\AlON}{$\mathrm{Al}_{2}\mathrm{O}_{3}(4\mathrm{N})$}  
\newcommand{\NaTmWO}{$\mathrm{Na}\mathrm{Tm}{(\mathrm{W}\mathrm{O}_{4})}_{2}$}  
\newcommand{\NaTmMoO}{$\mathrm{Na}\mathrm{Tm}{(\mathrm{Mo}\mathrm{O}_{4})}_{2}$}  

\newcommand{\Tm}{$^{169}\mathrm{Tm}$}  
\newcommand{\Fe}{$^{57}\mathrm{Fe}$}  
\newcommand{\Kr}{$^{83}\mathrm{Kr}$}  
\newcommand{\Li}{$^{7}\mathrm{Li}$}  
\newcommand{\Cs}{$^{137}\mathrm{Cs}$}  
\newcommand{\Am}{$^{241}\mathrm{Am}$}  
\newcommand{\Uf}{$^{235}\mathrm{U}$}  
\newcommand{\Ue}{$^{238}\mathrm{U}$}  

\newcommand{\fA}{$f_A$}  
\newcommand{\mA}{$m_A$}  
\newcommand{\gAN}{$g_{AN}$}  
\newcommand{\gAg}{$g_{A\gamma}$}  
\newcommand{\gAe}{$g_{Ae}$}  

\begin{document}

\begin{abstract}
	The {\Tm} nuclide has first nuclear level at $8.41$~keV with magnetic type transition to the ground state and, therefore, can be used as a target nucleus for the search of resonant absorption of solar axions.
	We plan to use a Tm-containing crystal of a garnet family {\TmAlO} as a bolometric detector in order to search for the excitation of the first nuclear level of {\Tm} via the resonant absorption of solar axions.
	With this perspective in mind, a sample of the {\TmAlO} crystal was grown and tested for its bolometric and optical properties.
	Measurements of chemical and/or radioactive contaminations were performed as well.
	In this paper we present the test results and estimate the requirements for a future low-background experimental setup.
\end{abstract}


\maketitle

\section{Introduction}
\label{sec:intro}

With the ongoing search for the dark matter, a hypothetical axion particle remains a valid candidate with robust theoretical motivation.
Having been introduced back in 1978, the hypothetical axion was initially supposed to solve the long-standing strong CP-problem in quantum chromodynamics (QCD).
The new pseudoscalar particle had to emerge after the newly introduced chiral symmetry had been spontaneously broken at some energy scale {\fA}, thus compensating the CP-violating term of QCD Lagrangian~\cite{Peccei1977,Weinberg1978,Wilczek1978}. The axion interactions with ordinary matter are described in terms of effective coupling constants {\gAN} (nucleons), {\gAg} (photons) and {\gAe} (leptons).
The values of these constants and axion mass {\mA} appear to be inversely proportional to the symmetry breaking scale {\fA} in all theoretical models.

The initial model of ``standard'' axion assumed the chiral symmetry breaking scale to be comparable to that of the electro-weak interactions ($f_A \approx 250\ \mathrm{GeV}$), thus making specific predictions regarding axion mass and coupling constants.
A series of experiments employing reactors, particle accelerators and artificial radioactive sources were carried out and eventually invalidated the existence of ``standard'' axion.
Nevertheless, there is no principal restrictions on the symmetry breaking scale \fA, which can be made arbitrary large, consequently reducing the expected axion mass and suppressing its interactions with ordinary matter.

Soon, after exclusion of the ``standard'' axion, new modified models were developed.
The axions described in these models were nicknamed ``invisible'', due to their weak coupling with ordinary matter~\cite{Kim1979,Shifman1980,Zhitnitski1980,Dine1981}.
Moreover, it turned out that such a particle fits the criteria for a potential dark matter constituent.
Thus, the search for axions and axion-like particles is considered an extremely important task.

With the intensely ongoing experimental searches for the dark matter particles there is a constant demand for advancement in low-background techniques as the experiments keep excluding new regions of the parameter space.
This paper reports the results of various measurements aimed to study the properties of {\TmAlO}, a thulium containing crystal of the garnet family, which can be used to detect the resonant absorption of solar axions~\cite{Derbin2007,Derbin2009,Derbin2011}.

\section{Thulium as a target material}
\label{sec:tm_target}

The majority of the axion experiments can be addressed to one of two major groups: solar axion searches and relic axion searches.
Several processes could be responsible for axion production inside the solar core; at present, the main efforts are focused on searching the conversion of solar axions to photons in a macroscopic laboratory magnetic field (CAST~\cite{Arik2009}, IAXO~\cite{Armengaud2014}).
The spectrum of solar axions, similarly to the spectrum of solar neutrinos, contains a continuous part with average energy about $4$~keV, produced by Primakoff effect, Compton-like processes and bremsstrahlung, and several monochromatic lines, associated with the emission of axions in nuclear transitions of magnetic type.

\begin{figure}[t]
	\includegraphics[width=\linewidth,trim=50 50 50 50]{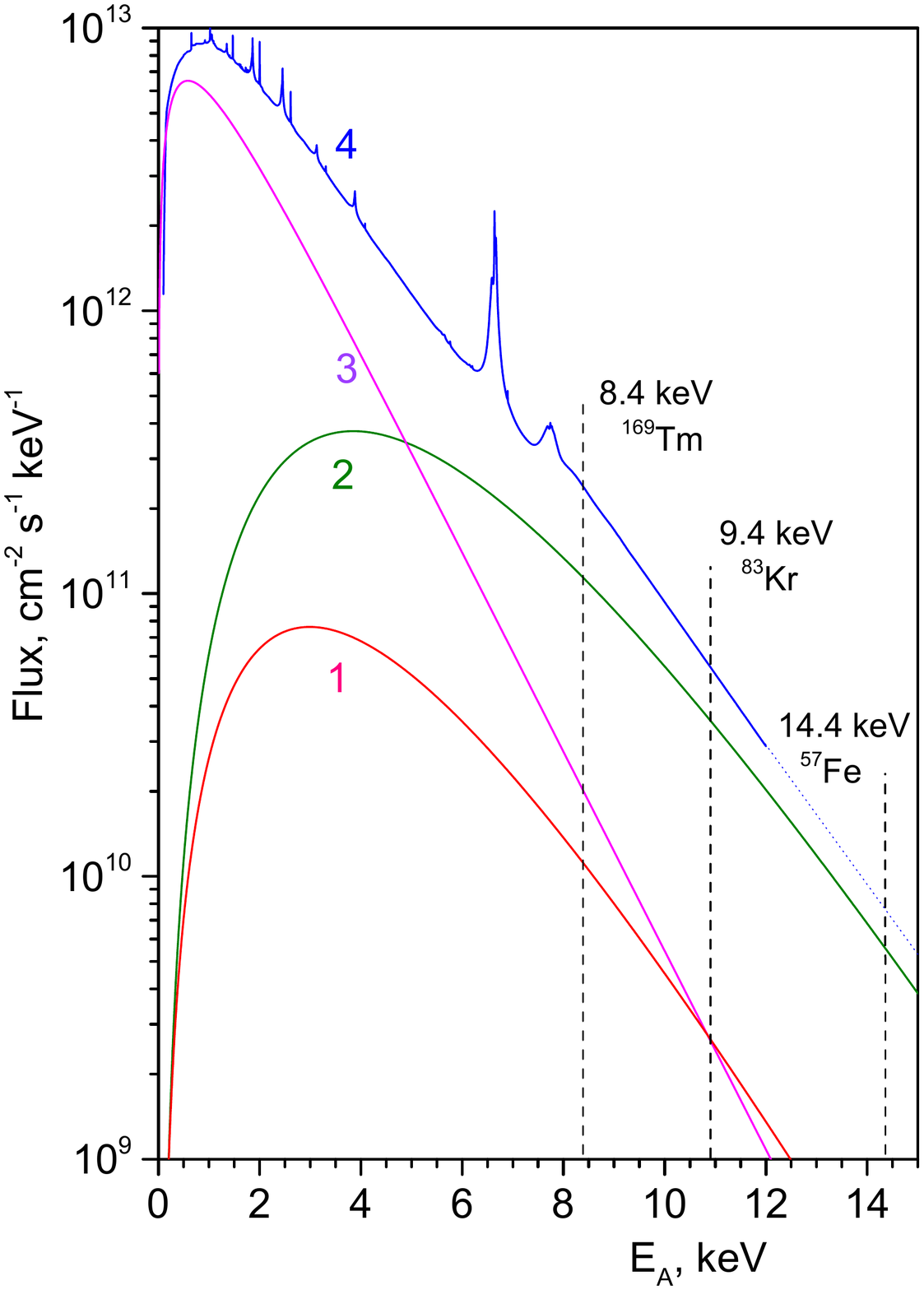}
	\caption{Solar axion spectra for Primakoff~(1), Compton-like processes~(2) and bremsstrahlung~(3), calculated for $g_{A\gamma} = 10^{-10}\ \mathrm{GeV^{-1}}$ and $g_{Ae} = 10^{-11}$, correspondingly. The total axion spectrum from the axion-electron coupling is represented by line~(4)~\cite{Redondo2013}.}\label{Fig:solar_flux}
\end{figure}

Among the possible axion interactions with ordinary matter, theoretical models describe the resonant absorption of an axion by atomic nuclei via the axion-nucleon ({\gAN}) interaction~\cite{Haxton1991,Avignone1988}.
The relatively high cross-section of this process allows achieving competitive sensitivity levels on a small scale setup.
The search for resonant absorption of monochromatic axions emitted by {\Fe}, {\Li} and {\Kr} nuclei at the Sun was proposed to be carried out in~\cite{Moriyama1995,Krcmar2001,Jakovcic2004}, respectively.

The use of {\Tm} low-lying nuclear level to search for axions with a continuous spectrum was proposed in~\cite{Derbin2007}. Searches for Primakoff, Compton and bremsstrahlung axions using {\Tm} were performed in~\cite{Derbin2009,Derbin2011} via the ``target-detector'' scheme.
A significant advantage of experiment with {\Tm} target is that the probability of axion emission/absorption in $8.41$~keV M1 transition depends weakly on the actual values of two poorly constrained QCD parameters ($S$ and $z$) as opposed to {\Fe} and {\Kr} nuclei, for which the probability can vanish in some cases~\cite{Derbin2009}.
The sensitivity of the experiments is limited by low detection efficiency, which can be significantly increased by introducing the {\Tm} target inside the sensitive volume of the detector.

A low-background setup equipped with a cryogenic detector constituted by a target containing {\Tm} allows one to test the resonant absorption of axions with a detection efficiency close to $100\%$ and a strong suppression of the possible background induced by photons coming from the excited atomic levels.
The latter consideration is critical for {\Tm}, since it has several energies of characteristic X-rays very close to $8.41$~keV $\gamma$-line~\cite{Derbin2009}.
The first attempts to employ Tm-containing crystals {\NaTmWO} and {\NaTmMoO} as a bolometer detector were undertaken in~\cite{Derbin2015}.

In this work we present the first experimental results obtained using a small crystal of {\TmAlO} with size of $(\sim10\times10\times10)$~mm$^3$ and weight of $8.18$~g.
The obtained results prove the feasibility of using the given material in a cryogenic calorimeter detector module and can be used to estimate the specifications for a future full-scale installation.

\section{Crystal growth characterization}
\label{sec:crystal}

\subsection{Crystal growth and sample preparation}
The {\TmAlO} crystal was fabricated at Kurchatov Institute in Moscow, Russia.
The crystal boule was grown from iridium crucible (diameter $40$~mm, height $40$~mm) by conventional Czochralski technique with RF-heating.

The original raw oxides {\YON}, {\TmON} and {\AlON} were annealed at $900^{\circ}$~C in order to remove moisture.
Afterwards, the ingredients were thoroughly mixed in stoichiometric proportion, pressed into tablets, sintered and placed into the crucible for crystal growth.
High purity zirconium oxide ($4$N) was used as thermal insulation to maintain the required temperature conditions inside the system.

The process was carried out in the nitrogen atmosphere with $0.1$~\% admixture of $\mathrm{O}_2$.
The YAG crystal oriented along $\langle 100 \rangle$ axis was used as crystallization seed.
The crystal was grown setting a pulling rate of $2$~mm/hour and rotation speed of $20$~rpm.
Such growth conditions provide the formation of the optimal convex shape of  the solid-liquid interface.

As a result of the growth process we obtained a $16$~mm diameter and $40$~mm height boule. 
This boule was then used to fabricate two crystal samples.
Sample~$\#1$ was cut from the boule tail, therefore the bottom surface has an irregular shape.
Overall, this sample measures $(8\times 10\times 10)$~mm$^3$ and weighs $5.5$~g.
The bottom surface of this sample appeared to be covered in small amount of iridium from crucible, while the rest of the surfaces were clear.
This sample was used to perform radiopurity and optical measurements.

Sample~$\#2$ was produced from the top part of the boule and has cubic shape with plain cuts on all surfaces.
Its dimensions are $10\times 10\times 10$~mm$^3$ with an $8.18$~g mass.
This sample was used in a cryogenic measurement to test the bolometric properties of the {\Tm}-containing garnet.

\subsection{Optical properties}
\begin{figure}[t]
	\includegraphics[width=\linewidth,trim=50 50 50 50]{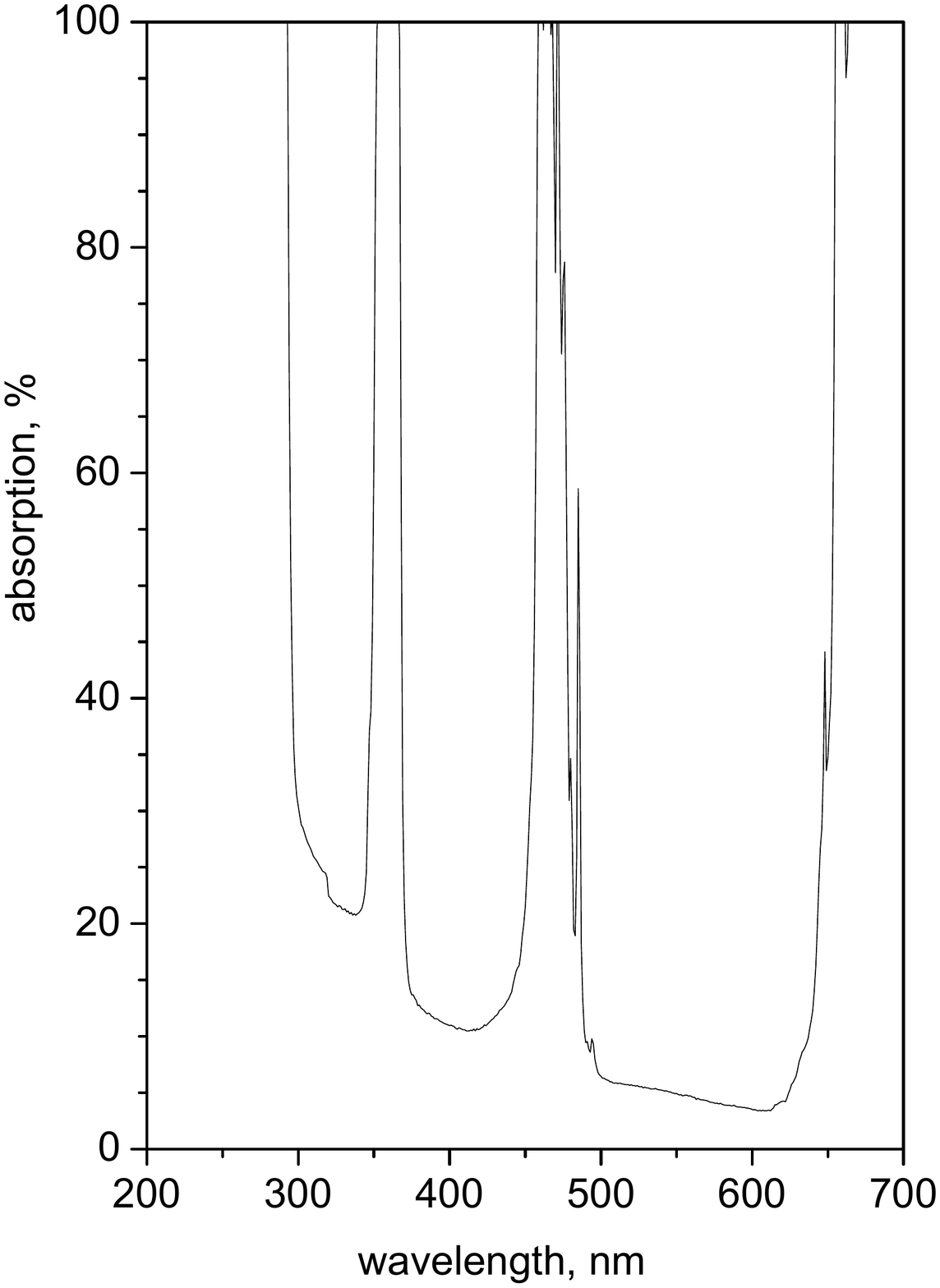}
	\caption{Absorption spectrum of the crystal sample~$\#1$ for the optical band.}\label{Fig:absorp}
\end{figure}

The absorption and transparency spectra of sample~$\#1$ were obtained in LNGS (Italy) using the UV-VIS spectrophotometer within $200-700$~nm wavelength range.
The normalized absorption spectrum is presented in Fig.~\ref{Fig:absorp}. The spectrum contains two distinct absorption bands, approximately at $350 - 367$~nm and $457 - 477$~nm.
Absorption bands approximately correspond to those associated with {\Tm} ions at $460$~nm and $681$~nm, reported in earlier research of {\Tm}-doped YAG~\cite{Korner2018,Gruber1989}.
Due to the presence of wide absorption bands within the visible range the use of the crystal as a scintillator would seem ineffective and for now no further investigation of scintillating properties was performed.

\subsection{Low-background spectrometry}
The level of internal radioactive contamination of sample~$\#1$ the {\TmAlO} crystal with respect to uranium and thorium natural decay chains, and, in particular, to their daughter nuclides, was investigated by means of $\gamma$-ray spectroscopy with an ultra-low background high purity germanium (ULB-HPGe) detector.
\begin{figure}[t]
	\includegraphics[width=\linewidth,trim=50 50 50 50]{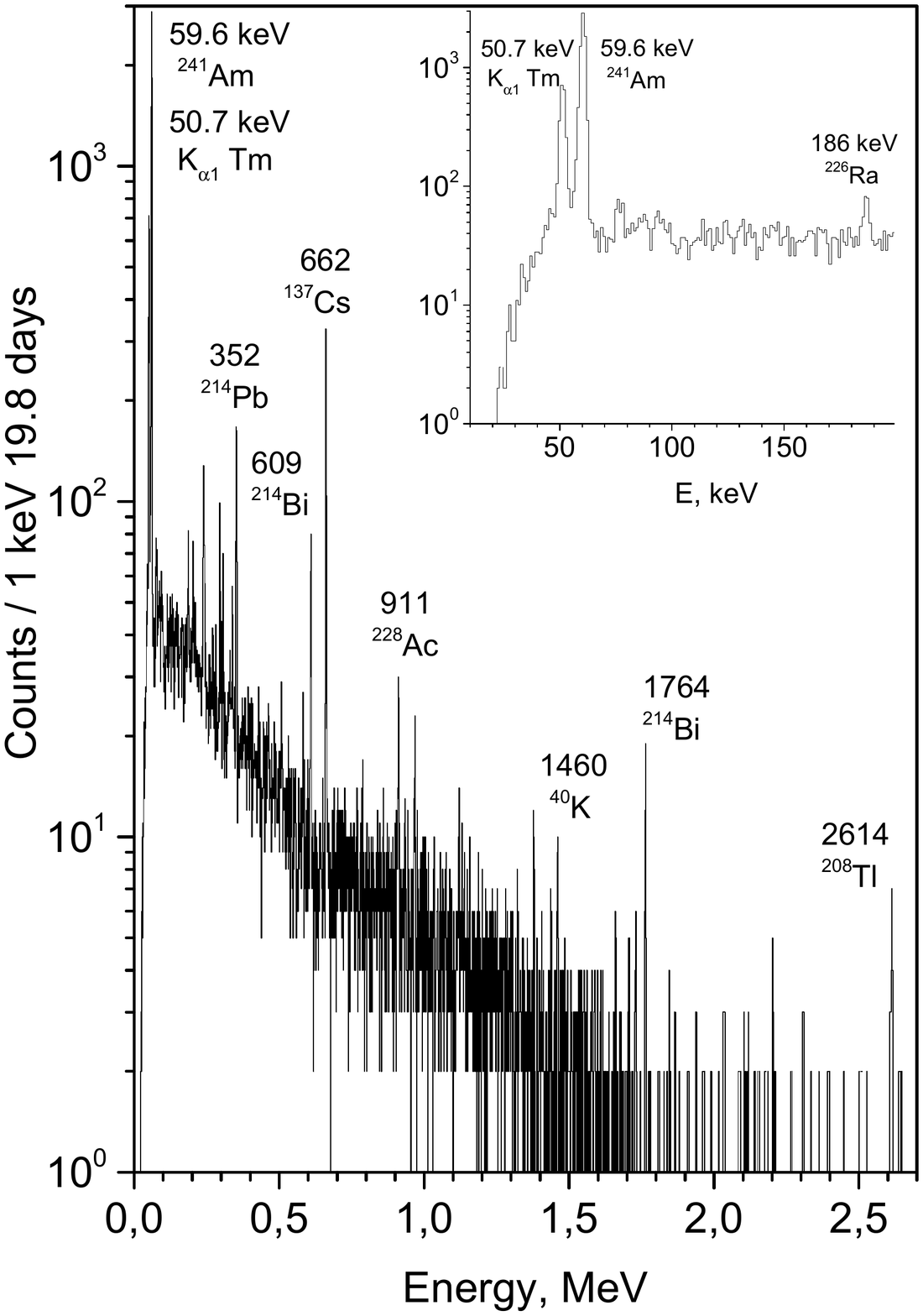}
	\caption{Energy spectrum of the crystal sample~$\#1$ obtained with Ge detector. The inset shows low-energy part of the spectrum.}\label{Fig:spec_Ge}
\end{figure}
The measurements were carried out in the STELLA (SubTerranean Low Level Assay) facility in the Gran Sasso National laboratories of the INFN in Assergi, Italy, which provides an average shielding of about $3600$~m.w.e.
Details about this facility can be found in~\cite{Arpesella1996,Laubenstein2017,Budjas2008,Hampel2008}.
The {\TmAlO} crystal sample~$\#1$ was placed on a well-type ULB-HPGe detector with an active volume of about $160$~cm$^3$.
This detector has a rather thin aluminum window of $0.75$~mm thickness and has an optimized design for high counting efficiency of small samples in a wide energy range.
The germanium detector is surrounded by low-radioactivity lead ($\approx 15$~cm), copper ($\approx 5$~cm) and special lead with low content in $^{210}\mathrm{Pb}$ ($\approx 5$~cm).
Finally, the shielding and the detector are housed in a polymethyl metacrylate box that is continuously flushed with pure nitrogen in order to suppress the amount of radon in the vicinity of the detector.

The energy resolution can be approximated in the energy region of $239 - 2615$~keV by the function:
\begin{equation}
	\mathrm{FWHM} = {(1.41(4) + 0.00197(4) \times E)}^{1/2}, 
\end{equation}
where $E$ is the energy of detected $gamma$-ray in~keV.
For instance, the FWHM at $1332.5$~keV gamma line of $^{60}$Co is $2.0$~keV.

The data with the {\TmAlO} crystal sample~$\#1$ was taken over $476.63$~hours ($19.86$~days), while the background spectrum was taken over $674.26$~hours ($28.09$~days).
The energy spectra of the {\TmAlO} crystal sample normalized to the time of measurement is presented in Fig.~\ref{Fig:spec_Ge}.

The specific activities of the isotopes were calculated using the formula:
\begin{equation}
	A = (S_s/t_s - S_b/t_b) / (y \cdot \eta \cdot m),
\end{equation}
where $S_s(S_b)$ is the area of a peak in the sample (background) spectrum, $t_s(t_b)$ is the time of the sample (background) measurement, $y$ is the yield of the corresponding $\gamma$-line, $\eta$ is the efficiency of the full peak detection and $m$ is the mass of the sample. 
The efficiencies for the full-energy absorption peaks used for the quantitative analysis were obtained through a Monte-Carlo simulation (code MaGe), based on the GEANT4 software package~\cite{Boswell2011}. 
The values of the limits were obtained using the procedure presented~\cite{Heisel2009}.
The nuclides and their activities found in the {\TmAlO} crystal samples are shown in Table~\ref{Tab:contamination}.

\begin{table}[t]
	\centering
	\begin{tabular}{rrl}
		\toprule
		Chain		&Nuclide	&Activity [Bq/kg]\\
		\midrule
		$^{232}$Th	&$^{228}$Ra	&$0.27 \pm 0.04$\\
				&$^{228}$Th	&$0.22 \pm 0.03$\\
		\midrule
		$^{238}$U	&$^{226}$Ra	&$0.45 \pm 0.03$\\
				&$^{234}$Th	&$2.5 \pm 0.9$\\
				&$^{234m}$Pa	&$\le 2.3$\\
				&$^{210}$Pb	&$4 \pm 1$\\
		\midrule
		$^{235}$U	&$^{235}$U	&$0.11 \pm 0.02$\\
		\midrule
		---		&$^{40}$K	&$\le 0.36$\\
				&$^{60}$Co	&$\le 0.020$\\
				&$^{241}$Am	&$94 \pm 9$\\
				&$^{137}$Cs	&$0.85 \pm 0.09$\\
				&$^{176}$Lu	&$0.09 \pm 0.01$\\
				&$^{138}$La	&$0.03 \pm 0.01$\\
		\bottomrule
	\end{tabular}
	\caption{The concentration of radionuclides (in~Bq/kg) in the {\TmAlO} crystal sample $\#1$, obtained by ULB-HPGe measurements.
	The upper limits are given with $90$\%~C.L., and the expanded standard uncertainties with $k = 1$.}\label{Tab:contamination}
\end{table}

\noindent The measurements have shown a significant contamination of the crystal sample~$\#1$ with {\Am} and {\Cs} isotopes.
The intensity of the $59.6$~keV {\Am} peak was $370$~events/day, which corresponds to the {\Am} $\alpha$-activity of $\sim 900$~decays/day.
{\Am} presence within the sample has no clear explanation at the moment.
One can speculate that iridium crucible used for the crystal growth had been previously exposed to {\Am}-containing material.
Sample~$\#1$ is also contaminated by La and Lu nuclides, which can be explained by chemical affinity of Tm and Lu/La.
Thulium does not occur in the nature in a free state, while it is commercially produced from minerals containing rare earth elements of the lutetium subgroup (from Gd to Lu).
Finally, sample~$\#1$ contains nuclides from U/Th decay chains.
The ratio of {\Uf}/{\Ue} is in agreement with natural abundance of uranium isotopes.
Secular equilibrium appears to be broken, although this is a typical occurrence in inorganic crystals~\cite{Danevich2003,Angloher2004,Zdesenko2005}.

It has been demonstrated before for various compounds, that the majority of impurities accumulate at the end of the crystal boule due to the segregation effect~\cite{Danevich2011}.
Thus, one could naturally expect that sample~$\#1$ from the bottom of the boule should be less radiopure with respect to the sample~$\#2$ from the top of the boule.

\subsection{Mass-spectrometry}
A general contamination screening for the wide range of elements was performed via the high-resolution inductively coupled plasma mass-spectrometry (HR-ICP-MS).
The measurement was performed by the ``Thermo Fisher Scientific ELEMENT2'' spectrometer located at the Gran Sasso National laboratories.
The material for measurements was obtained from the crystal sample~$\#2$ in form of crystal particulates, which were then dissolved in an acid solution and diluted for the measurement.
A semi-quantitative analysis was performed, i.~e. the instrument was calibrated via a single reference standard solution of thorium and uranium.

While the chemical purity was analyzed with respect to a wide range of elements, we would like to stress the attention on some of them, which have critical importance for crystal quality, or as elements that affects crystal radiopurity.
For the most of the elements only the upper limits on concentration were obtained.
The obtained concentrations for various elements are presented in Table~\ref{Tab:icp-ms}.

\begin{table}[t]
	\centering
	\begin{tabular}{rl|rl|rl}
		\toprule
		Elem.&C~[ppm]&		Elem.&C~[ppm]&	Elem.&C~[ppm]\\
		\midrule
		K&	$\leq 21.4$&	Y&	$1357.1$&	Dy&	$1.6$\\
		Ca&	$15.7$&			Zr&	$70.7$&		Er& $6.7$\\
		Cr&	$\leq 3.0$&		Mo& $24.3$&		Yb& $4.8$\\
		Mn&	$\leq 0.7$&		I&	$71.4$&		Lu& $3.4$\\
		Fe&	$\leq 21.4$&	La&	$2.5$&		Hf& $7.1$\\
		Co&	$\leq 0.4$&		Ce&	$3.0$&		Ir& $6.9$\\
		Ni&	$\leq 3.6$&		Nd&	$4.4$&		Tl&	$2.6$\\
		Cu&	$\leq 0.7$&		Sm&	$1.2$&		Pb&	$78.6$\\
		Zn&	$0.7$&			Eu&	$0.4$&		Bi&	$0.4$\\
		Ga&	$1.9$&			Gd&	$11.4$&		Th&	$\leq 0.1$\\
		Br&	$7.9$&			Tb&	$0.3$&		U&	$0.1$\\
		\bottomrule
	\end{tabular}
	\caption{The element concentrations (C) in parts per million (ppm) units reported by the ICP MS study of the crystal sample~$\#2$.}\label{Tab:icp-ms}
\end{table}

One could see that transition elements of Fe group that have a huge impact of optical properties of any crystal are practically absent.
This confirms the fact that the green tint of {\TmAlO} crystal is caused by Tm ion properties rather than by the presence of such impurities.

The evidence of Ir on the level of $7$~ppm is caused by the high rate of material evaporation from the iridium crucible caused by exposure to high temperatures during the crystal growth process.
The use of inert atmosphere with a small admixture of oxygen during the crystal production may reduce such evaporation, though it cannot be eliminated completely.
One should notice that such concentration of Ir is observed inside the inner crystal volume, while the macroscopic iridium particulates cover one of the surfaces of the crystal sample~$\#2$.
Other elements of Pt-group are excluded at the concentration levels of less than $0.5$~ppm.

Apparently, the whole range of rare earth elements proved to be present in the final crystal material with concentrations of up to tens of ppm.
The elements of the Gb sub-group are present in larger amount with respect to the Ce sub-group, due to the the chemical affinity of thulium with the given elements.
The presence of Sm and Gd that have $\alpha$-decaying isotopes with relatively short half-lives may become the source of irreducible background.
If the particle interactions are being registered solely via the heat channel such background events would be indistinguishable from the pulses we look for.
Thus, for the high sensitivity experiment the concentration of rare earth elements inside the Tm-containing crystal should be thoroughly minimized.
The significant yttrium concentration of more than $1300$~ppm demonstrate that the declared purity grade (5N) of the Tm oxide powder should be double checked by independent measurements, and a reliable producer of thulium oxide should be selected accordingly.

Presence of the elements like Zn, Ga, Zr, Br, Mo, I, Hf, Tl, Pb, Bi could be possibly explained by instrumental contamination, caused by evaporation from the walls of the crucible, in case those elements were involved in previous crystal growths.
Therefore, in order to achieve the high purity Tm-containing crystal one should use a freshly produced iridium crucible and thermal shield.

With respect to the radioactive elements, the measurements showed high concentration of uranium ($0.1$~ppm) due to chemical affinity with rare earth elements, and the limit was set on the presence of thorium ($\leq 0.1$~ppm).
The limit on potassium concentration was found to be $\leq 22$~ppm.
The exposure of crystal material to high temperatures during the growth helps to eliminate potassium impurities from the compound due to its high volatility.

\section{Experimental set-up}
\label{sec:exp}

\begin{figure}[t]
    \includegraphics[width=\linewidth]{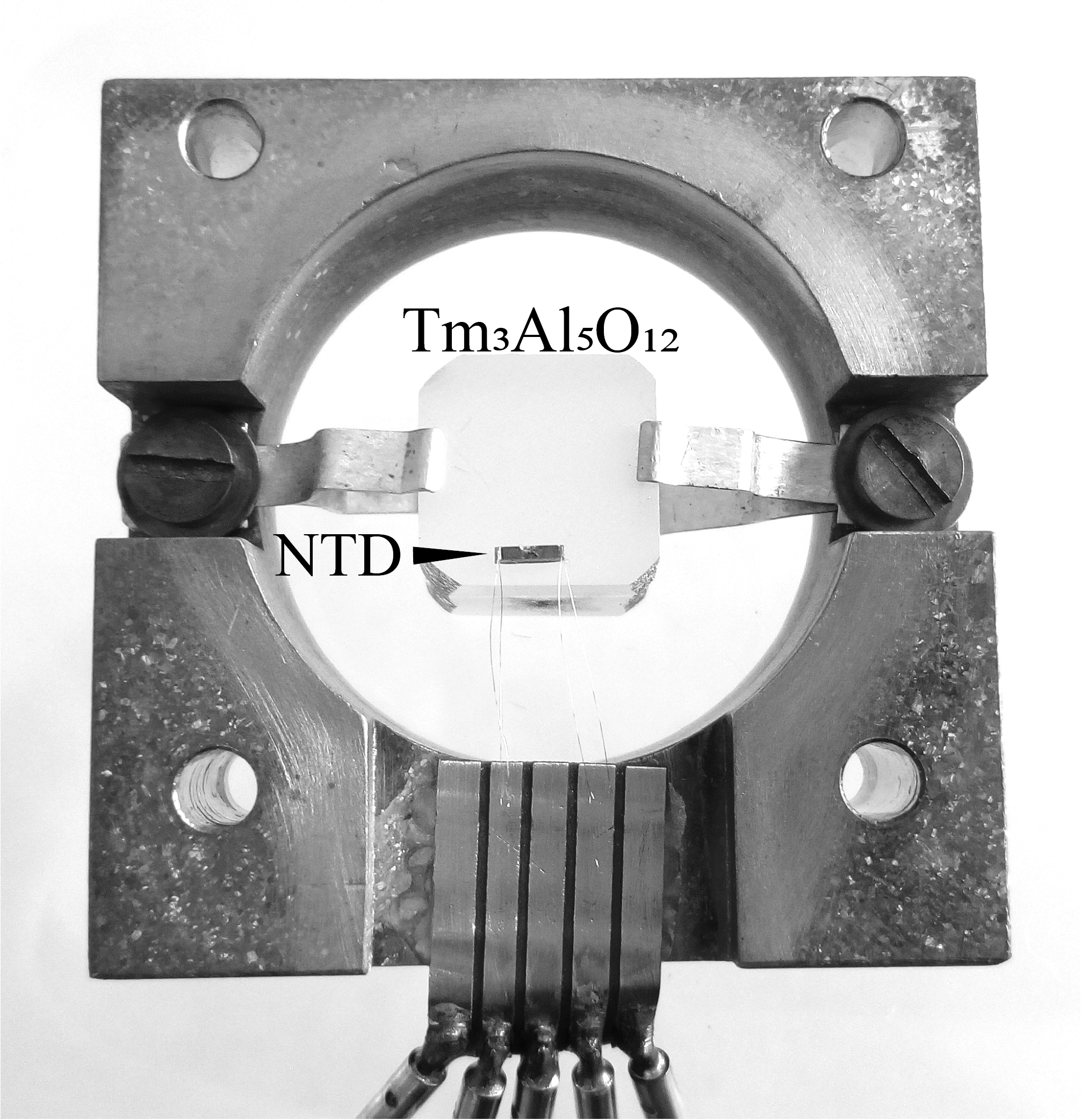}
    \caption{{\TmAlO} crystal sample~$\#2$ instrumented with a NTD inside the copper holder.
    Gold bond wires provide thermal and electrical connections to the bond pads glued on the copper holder.}
    \label{Fig:det_pic}
\end{figure}

The $8.8$~g {\TmAlO} crystal (sample~$\#2$) is used as main absorber of a cryogenic detector.
The absorber is held in position inside a copper holder by two pairs of bronze clamps.
One surface of the crystal is coupled to a Neutron Transmutation Doped (NTD) sensor~\cite{NTD} through a thin layer of epoxy~\footnote{GP 12 Allzweck-Epoxidkleber}; electrical and thermal connections are provided to the NTD through a pair of gold bond wires with a $25$~$\mu$m diameter.
This detector (see Fig.~\ref{Fig:det_pic}) was operated at the Max-Planck-Institute for Physics (MPI) in Munich, Germany, inside a dilution refrigerator~\footnote{Kelvinox400HA} in an above-ground laboratory without any shielding against the environmental and cosmogenic radiation.
The detector was mechanically and thermally connected to the coldest point of the dilution refrigerator: $\sim$10~mK was the lowest temperature reached during the measurement.
The temperature readout of the NTD is obtained measuring the voltage drop variations of the sensor with a differential voltage amplifier while applying a constant bias current through the NTD.
At cryogenic temperatures, a particle interacting inside the target crystal produces a thermal pulse that follows a well-tested model described in~\cite{tes}.
The pulse amplitude is proportional to the energy deposited in the absorber, so,with the help of calibrations sources, it is possible to accurately measure the energy spectrum of the particle interactions above a certain energy threshold $E_{r}$.

\begin{figure}[t]
    \centering
        \includegraphics[width=\linewidth]{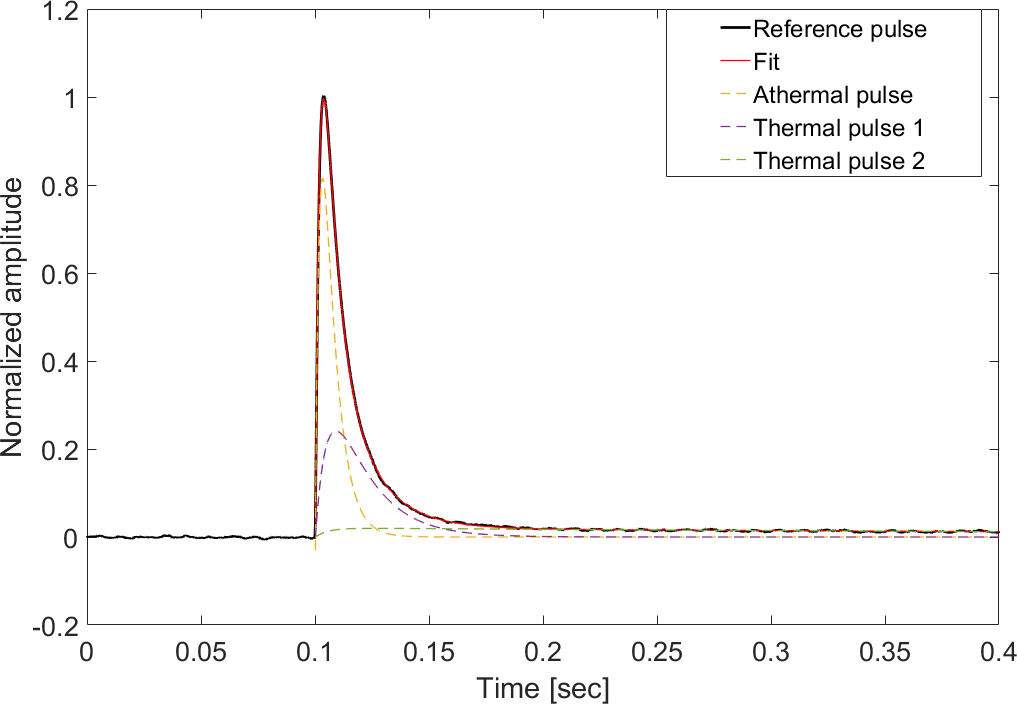}
        \caption{In solid black a typical particle interaction inside the {\TmAlO} crystal measured by the NTD.
        In solid red the parametric fit.
        In dashed yellow, dashed purple, dashed green the three exponential components: athermal, first thermal component, and second thermal component respectively.
        The second thermal component is not included in the physical model usually adopted and has a much longer lifetime than the other two exponential components.}\label{Fig:PulsefigV1}%
\end{figure}

From~\cite{tes} we expect a measured particle pulse to be a superposition of two exponential pulses.
In our case, however, we had to introduce a third exponential pulse in order to properly describe the pulse shape of a particle interaction taking place inside the {\TmAlO} crystal, as shown in figure~\ref{Fig:PulsefigV1}.
We attribute this third exponential pulse to a second thermal component, since a similar effect has already been seen in previous works with cryogenic detectors~\cite{cosinus}.
The interpretation of this second thermal component is not straightforward and si under investigation.
The fit of a typical particle pulse measured by our detector can give useful information about the exponential components such as the life time of non-thermal phonons ($\tau_{n}$), the intrinsic thermal relaxation time constant of the thermometer ($\tau_{in}$), and the thermal relaxation time constants of the absorber ($\tau_{t1}$, $\tau_{t2}$).
The result of the fit in this case leads to $\tau_{n}$=5.5~ms, $\tau_{in}$=1.9~ms, $\tau_{t1}$=15~ms, and $\tau_{t2}$=560~ms.
It is worth noticing that the additional thermal component we introduced ($\tau_{t2}$) appears to live in a long time scale: a thermal component with a long lifetime might affect the accuracy of the energy reconstruction of particle interactions, especially in the presence of a high rate.
Thus, it might be even more beneficial than usual to operate this detector in a low background environment.

\section{Results}
\label{sec:results}

\begin{figure}[t]
    \centering
    \includegraphics[width=\linewidth,trim=50 50 50 50]{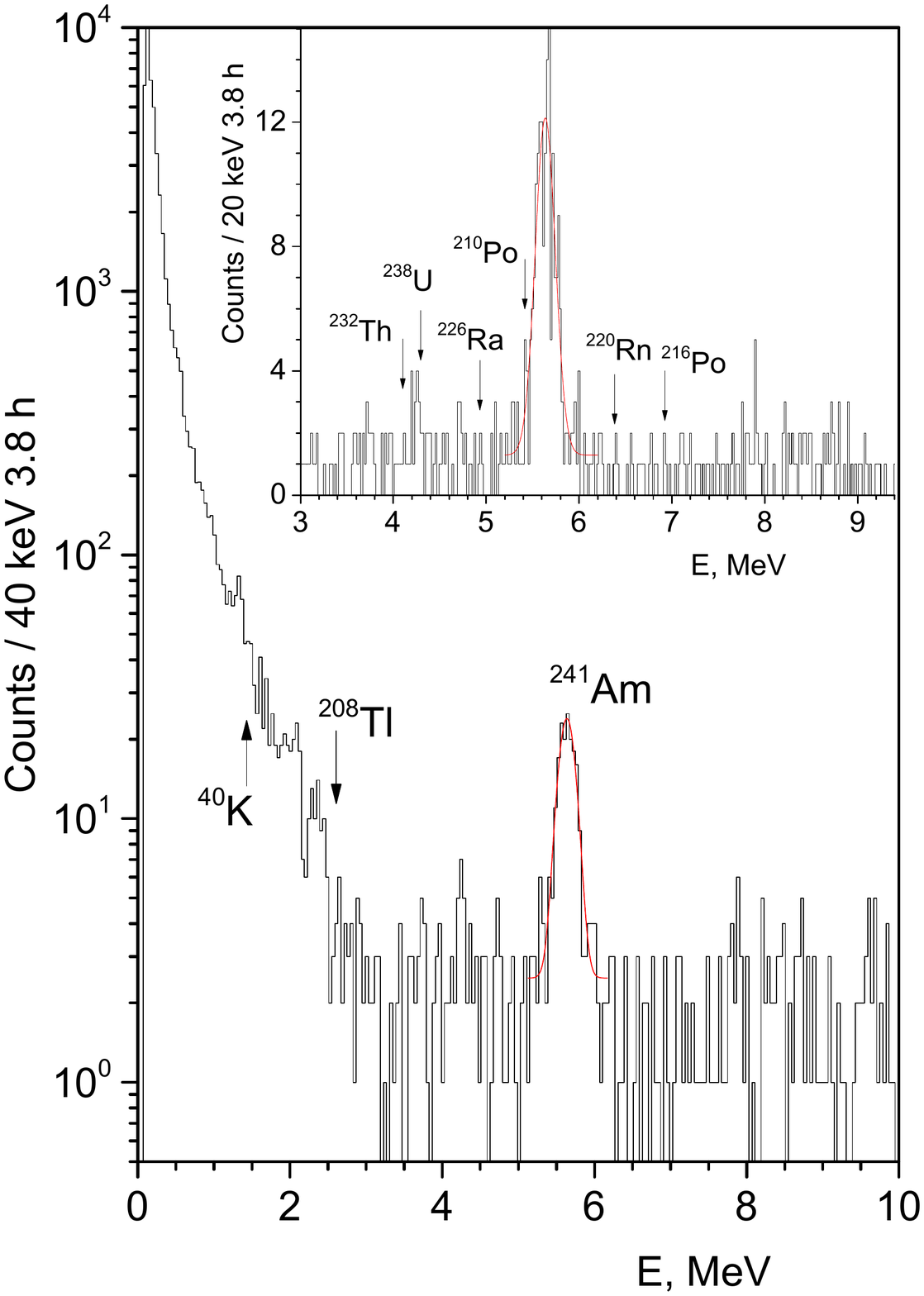}
    \caption{Energy spectrum measured by the NTD for a $1.3$~g$\cdot$day exposure.
    Ths inset shows the energy interval where $\alpha$-peaks of {\Ue} and $^{232}$Th chains should be manifested.
    The energy calibration was obtained using the mean value of a broad peak that we attribute to the $5.637$~MeV alpha decay of {\Am}, the main contaminant present in the crystal.}\label{Fig:spec}
\end{figure}

This measurement convincingly shows that it is feasible to operate a bolometric detector which employs {\TmAlO} as the absorber.
This is, of course, the necessary condition for a cryogenic detector with a new target material.
This, of course, is the first goal to achieve before planning a cryogenic experiment based on a given target material.
{\TmAlO} is particularly attracting for solar axions search, on the condition that it is possible to obtain an energy threshold $\leq~8$~keV.
Thus, this measurement was also useful to understand how close we are to this design goal with the technology employed.
In this first test, we have recorded 3.8 hours of data in a continuous single run.
For the analysis we have considered the whole dataset without the application of any cut, in order to preserve the 1.3~g$\cdot$day exposure.
Due to the high content of long living radioactive elements it is possible to immediately identify characteristic features in the energy spectrum, despite the low exposure.
A broad peak evidently appears in the spectrum shown in figure~\ref{Fig:spec}: this peak can be ascribed to the alpha decay of $^{241}$Am (5.637~MeV Q-value~\cite{nuclear}), the highest contaminant present in the sample.\\ Consequently, it is possible to use this peak to calibrate the detector response.

The results of the peak fitting via the gaussian are given in Fig.~\ref{Fig:spec}.
The peak position corresponds to $Q_{\alpha} = 5.637$~MeV and was used for energy scale calibration.
The peak width was determined to be $\sigma = 110$~keV, although this evaluation might be affected by the complex structure of the peak, which is formed by the escape probabilities of $59.6$~keV and $43.4$~keV $\gamma$-quanta and characteristic X-rays.
Number of evens within the peak amounts to $155 \pm 10$ events and corresponds to {\Am} activity of $1.4$~Bq/kg.

The rest of the spectrum does not contain any prominent $\alpha$-peaks.
In order to estimate the activity of various nuclides belonging to $\mathrm{U}$ and $\mathrm{Th}$ decay chains we used the events within $3\sigma = \pm 330$~keV interval from the the $Q_{\alpha}$ value.
The spectrum was fitted by gaussian with fixed parameters of $x_0 = Q_{\alpha}$ and $\sigma = 110$~keV, while the peak area and the background constant remained free.
The obtained upper limits for activity of some isotopes are given in Tab.~\ref{Tab:activity}.
Mixed $\alpha-\beta$ decays from $^{214}$Bi-$^{214}$Po and $^{212}$Bi-$^{212}$Po decay chains appear in the spectrum as single events: unfortunately, time-coincidence techniques cannot be employed due to the relatively slow pulses of the bolometric detectors. 

The measured activities are considerably smaller than the ones measured in Gran Sasso, pointing towards a strong segregation of the contaminants during the crystal growth.
To estimate the background introduced by these contaminants at the energy of interest ($\sim$8~keV) a further investigation on the internal contamination of this crystal is needed.

\begin{table}
    \centering
    \begin{tabular}{rl|rl}
        \toprule
        Isotope&A~[Bq/kg]&    Isotope&A~[Bq/kg]\\
        \midrule
        $^{238}$U&  $\leq 0.28$&    $^{232}$Th& $\leq 0.16$\\
        $^{235}$U&  $\leq 0.24$&    $^{230}$Th& $\leq 0.20$\\
        $^{234}$U&  $\leq 0.14$&    $^{210}$Po& $\leq 0.33$\\
        $^{226}$Ru& $\leq 0.12$&    $^{241}$Am& $1.4 \pm 0.1$\\
        $^{220}$Ru& $\leq 0.10$&   ---&        ---\\
        \bottomrule
    \end{tabular}
    \caption{Upper limits and values of activities (A) in Bq/kg of U and Th chain contaminants, determined by the bolometric measurement of crystal sample~$\#2$ ($90$\%~C.L.)}\label{Tab:activity}
\end{table}

\section{Conclusions}
\label{sec:conclusions}

The performed measurements have proven for the first time the feasibility of operating a thulium-containing crystal as a cryogenic bolometer.
The obtained baseline resolution amounted to $22.75 \pm 0.65$~keV, which translates to energy threshold $E_{th}$ of about $170$~keV.
The phonon events pulse shape was well-fitted via the four time constraints fit, where the longest component ($560$~ms).

While the principle possibility of operating a thulium-based material in a cryogenic set-up has been confirmed, it is clear that further optimization is needed in order to achieve the required sensitivity for the solar axion search.
In order to obtain physical results we need an improvement of more than one order of magnitude for the energy threshold, which might be in reach using a CRESST-like detector with a TES as thermal sensor: CRESST has already demonstrated the ability to obtain outstanding energy thresholds employing small crystals ($30.1$~eV with a $24$~g CaWO$_4$ crystal~\cite{Abdelhameed2019} and $19.7$~eV with a $0.5$~g Al$_2$O$_3$ crystal~\cite{detect_concept}).

The screening of Tm-based compound revealed significant contamination by U/Th decay chains at the level of few Bq/kg.
The crystal also contains isotope {\Cs} with activity of about $1$~Bq/kg, as well as radioactive nuclides $^{176}$Lu and $^{138}$La that accompany rare earth metals in raw materials ({\TmON} and {\YON} powders).
The presence of {\Am} a:t the level of $100$~Bq/kg together with {\Cs} in the bulk of the crystal remains poorly understood, not excluding the possible contamination from growth equipment or raw materials.

In order to produce the low-background Tm-containing crystal that will meet the requirements for chemical and radioactive purity one should exercise additional precautions during the crystal production and handling, including pre-growth purification of {\TmON} and {\YON} powders against U/Th chain nuclides and thorough screening of the growth equipment to be used.
Further investigations of the application of the Tm-containing crystal as a cryogenic bolometer are ongoing.

\section{Acknowledgements}

This work was supported by the Russian Foundation of Basic Research (grants 17-02-00305A, 16-29-13014ofi-m and 19-02-00097A).

\section*{References}

\bibliographystyle{elsarticle-num}
\bibliography{references}

\end{document}